\documentclass[epj,final]{svjour}

\usepackage{latexsym}
\usepackage{url}
\usepackage{amsfonts}
\usepackage{amsmath}\usepackage{amssymb}

\RequirePackage{graphicx}
\usepackage{epstopdf}
\begin{document}
\title{Non-random structures in universal compression and the Fermi
  paradox }
\author{A.V. Gurzadyan\inst{1} \and A.E. Allahverdyan\inst{2}
}                     
%
%
\institute{Russian-Armenian (Slavonic) University,
  Emin Street 123, Yerevan 0051, Armenia \and 
Yerevan Physics Institute, Alikhanian Brothers Street 2,
  Yerevan 0036, Armenia}
\date{Received: date / Revised version: date}
%

\abstract{
We study the hypothesis of information panspermia assigned recently
  among possible solutions of the Fermi paradox (``where are the
  aliens?''). It suggests that the expenses of alien signaling can be
  significantly reduced, if their messages contain compressed
  information. To this end we consider universal compression and
  decoding mechanisms (e.g. the Lempel-Ziv-Welch algorithm) that can
  reveal non-random structures in compressed bit strings. The
  efficiency of Kolmogorov stochasticity parameter for detection of
  non-randomness is illustrated, along with the Zipf's law. The universality
  of these methods, i.e. independence on data details, can be
  principal in searching for intelligent messages. 
} 

\maketitle
\section*{Introduction}

The Fermi paradox \cite{webb} outlines the puzzling silence of the
universe regarding intelligent life. It is a widely discussed topic
that connects to various disciplines, from astrophysics to
linguistics. The concept of information panspermia introduced in
\cite{G} and listed in \cite{webb} as Solution 23 to the Fermi
paradox, implies the transmission of terrestrial life via compressed
strings. The idea is based on the estimation \cite{G} that Earth
organisms|including humans, and the entire terrestrial life up to
bacteria|have common parts in their genomes. Hence the entire genome
can be efficiently compressed|the theoretical, but uncomputable, lower
bound of the compression degree is given by the Kolmogorov complexity
\cite{K,cover}|and the resulting bit string can be transmitted to over
Galactic distances, e.g. by Arecibo type radio telescope.  After
decoding, signals can represent themselves as traveling life
streams. One can imagine that the terrestrial life is a result of such
a transmitted package.

The principal difficulty with checking this hypothesis is that the
decoding of such signals has to be based on criteria that differ from
those employed in intra-terrestrial communication. For instance, the
spectral (monochromaticity) and temporal (modulations) features are
likely to be not efficient for extra-terrestrial messages
\cite{webb}. Indeed, an efficient signal has to behave as white noise
by its statistical properties and hence it cannot be distinguished
from natural signals, except in the cases when the precise code is
known in advance \cite{shostak}. When searching for intelligent
signals one looks for ``artificial'' features such as
monochromaticity. The latter is however costly, especially at
semi-isotropic transmission.

Thus traveling bit strings require different strategies for
distinguishing them from known physical mechanisms and eventually for
their decoding. Our aim is to outline several strategies based on
universal information compression and decoding features.

{\bf Conjecture}: {\it Once intelligent bit string signals will be
  compressed, we have to look at and reveal universal features of
  information compression which are absent in compressed signals
  originated via natural physical mechanisms.}

Below we suggest schemes that possibly enable to reveal non-random
structures in bit strings. We focus on universal schemes, i.e. those
applying to the large classes of situations. This is consistent with
Minsky's methodology on the priority of universal concepts|in computer
science and (more generally) mathematics|when searching for
intelligent signals \cite{minsky}.

\section{Kolmogorov's complexity}
\label{ks}

We start with the most fundamental notion of compression. 

The Kolmogorov's complexity $K(x)$ of a bit string $x$ with length $n$
is defined to be the length (in bits) of the shortest program
that|starting from some fixed initial state|will generate $x$ and halt
\cite{K,cover}:
\begin{equation}
\label{kar}
K(x)={\rm min}\, l(x).
\end{equation}
$K(x)$ depends on a specific computer that executes candidate
programs. This dependence is weak in the following sense: there exists
a universal computer ${\cal U}$ such that for any other computer
${\cal C}$ \cite{K,cover}:
\begin{eqnarray}
  \label{eq:3}
K_{\cal
  U}(x)< K_{\cal C}(x)+C_{\cal C},  
\end{eqnarray}
where $C_{\cal C}$ depends only on ${\cal C}$ (it does not depend
on $x$), and $C_{\cal C}/n\to 0$ for $n\to\infty$. The validity of
(\ref{eq:3}) stems from the fact that any computer operates with
finite means, hence a program to be executed on ${\cal U}$ can be
represented as $\pi_{\cal C}p_{\cal C}$, where $\pi_{\cal C}$ is the
index (coordinate) of ${\cal C}$, while $p_{\cal C}$ is a program of
${\cal C}$. Since $\pi_{\cal C}$ is finite, we are back to
(\ref{eq:3}).

Note that the notion of the computation time is not involved in
definition (\ref{kar}).

Thus the Kolmogorov's complexity does provide a universal lower bound
for the length of possible compressions of $x$ that can recover $x$
uniquely. 

Eq.~(\ref{kar}) easily generalizes to the conditional complexity
$K(x|y)$, where during the (conditional) minimization all programs are
assumed to start with $y$, i.e. the latter is known. In particular,
one can consider $K(x|l(x))$: the conditional complexity given the
knowledge of the string length.

It is important that almost all long bit strings $x$ are
incompressible or {\it algorithmically random} \cite{K,cover}:
\begin{eqnarray}
  \label{eq:4}
K(x|l(x))\geq l(x),
\end{eqnarray}
i.e. their (conditional) complexity is not much shorter than their
length \cite{K,cover}. This follows from the fact that there are $2^n$
bit strings of length $n$, while the amount of shorter strings
(i.e. shorter programs) is much less, i.e. there are $2^{n-k}\ll 2^n$
bit strings of length $n-k$. It thus is contradictory to assume that
almost all strings of length $n$ can be generated by such programs.
An example of an arbitrary long string that can be generated by a
short program is the $n$-digit rational approximation to $\pi$ or to
$e$. 

On the other hand, we have
\begin{eqnarray}
  \label{eq:44}
K(x)< K(x|l(x))+\log_2^*[l(x)]+{\rm const},
\end{eqnarray}
where
\begin{eqnarray} 
\log_2^*[l(x)]\equiv\log_2[l(x)]+\log_2\log_2[l(x)]+...,
  \label{eq:444}
\end{eqnarray}
and iterated logarithms continue as far as the logarithm is
well-defined. Indeed, knowing $l(x)$ we can just order the computer to
print $x$; the knowledge of $l(x)$ is obligatory, since the computer
should know where to halt. This knowledge demands $\log_2[l(x)]$ bits,
while the knowledge of the latter string demands $\log_2\log_2[l(x)]$
bits {\it etc}.

However, it is impossible to prove that some single bit string of
length $n$ is not compressible, if $n$ is large enough.  Thus the
corresponding function is not computable \cite{cover,K}. To understand
this point order the strings in a lexicographic way (i.e. $0$ comes
before $1$, $00$ comes before $01$ {\it etc}) and try to look for the
first string of length $n$ whose complexity is not smaller than
$n$. If such a string can be found, then the very description ``first
string of length $n$ whose complexity is not smaller than $n$''
constitutes a rather short program for that string, hence this string
is compressible to a large extent. The resulting contradiction shows
that although most of strings are incompressible, one can never show
that an individual string is incompressible, because showing that
would immediately mean that the string is compressible. This
undecidability is the major hindrance for practical applications of
the Kolmogorov's complexity. Another implication of the undecidability
is that one can never compute after how many steps $n$ a given (even a
short) program is going to halt. For, if we can compute $n$ for any
such program, we shall sequentially check all short programs and
eventually determine that certain strings are not compressible.

Non-compressible strings hold all feasible (computable) features of
probability theory. The precise definition of randomness (which is
equivalent to the algorithmic randomness) was given by Martin-Lof
\cite{martin}. It does improve upon the the earlier concept of random
``kollektiv'' by von Mises, which does not support all features of the
probability theory (e.g. the law of the iterated logarithm)
\cite{martin}.

For an ergodic, finite-memory random process $(X_1,...,X_n)$, the
Kolmogorov's complexity of almost any realization $(x_1,...,x_n)$
converges for $n\to\infty$ to the entropy of the random process
\cite{cover}:
\begin{equation}
  \label{eq:2}
|H(X_1,...,X_n)-
C(x_1,...,x_n)|={\cal O}(\log_2 n),
\end{equation}
and
\begin{equation}
H(X_1,...,X_n)\equiv  -\sum_{x_1,...,x_n} P(x_1,...,x_n)\log_2 
P(x_1,...,x_n),\nonumber
\end{equation}
where $P$ is the probability. Note that 
\begin{eqnarray}
  \label{eq:9}
  H(X_1,...,X_n)={\cal O}(n), ~~~~
C(x_1,...,x_n)={\cal O}(n).
\end{eqnarray}

Eq.~(\ref{eq:2}) is consistent with the (first) Shannon's theorem that
determines $H(X_1,...,X_n)$ to be the minimal number of bits necessary
for describing (with a negligible error) realizations of the random
process \cite{cover}.

\section{Universal coding on the example of Lempel-Ziv-Welch algorithm}

Two related aspects of the Kolmogorov's complexity is that it does not
involve the computation time|hence the search for a shorter
description can take indefinitely long|and that it is not computable
\cite{K,cover}. It is useful for establishing bounds \cite{G}, but it
cannot be employed for quantifying the degree of
compression/complexity in practice.

At this point it is useful to employ the notion of a universal,
asymptotically optimal data-compression code that is to large extent
inspired by the notion of the Kolmogorov's complexity
\cite{cover}. Here asymptotically optimal means that when applied to
random, ergodic processes, the average code-length of the code
approaches for long sequences the entropy of the process, i.e. the
analogue of (\ref{eq:2}) holds. And universal means that for achieving
this optimal performance, the code need not be given the frequencies of
the symbols to be coded \cite{cover}. Normally, the
instantaneously-decodable (i.e. prefix-free) feature is also assumed
within the definition of universality. Recall that prefix-free codes
can be decoded on-line, i.e. without knowing the whole (long) message
\cite{cover}. (Otherwise, the knowledge of the space-symbol is
required for the decoder \cite{cover}.)

Recall that several standard examples of optimal data-compression
codes|e.g. the Huffman's code or the arithmetic code|do demand a
priori knowledge of symbol frequencies, since they operate by coding
more frequent symbols via shorter code-words (the same idea is
employed already in the Morse code) \cite{cover}.

The most known (but not the only) example of a universal,
asymptotically optimal data-compression code is the Lempel-Ziv-Welch
algorithm \cite{lz}; see \cite{cover} for a review.  The algorithm and
its modifications have been widely used in practice; although newer
schemes improve upon them, they provide a simple approach to
understanding universal data compression algorithms. The rough idea of
this method is as follows \cite{lz,cover}: the string to be coded is
parsed (e.g. by commas) into substrings that are selected according to
the following criteria. After each comma one selects the shortest
substring that has not occurred before. Hence the prefix of this
substring did occur before, and the parsed substring can be coded via
the coordinate of this occurrence and the last symbol (bit) of the
substring.

The Lempel-Ziv-Welch algorithm presents an example of simple and
robust scheme that is relatively easy to uncover. For a given string
the code-length of its Lempel-Ziv-Welch compression provides an
estimate for the algorithmic complexity. This Lempel-Ziv complexity
\cite{lz_complexity} is also widely employed in practice, in
particular in biomedical research \cite{aboy}.

For our purposes it is important to stress that the Lempel-Ziv
complexity can be employed for detecting non-random structure inherent
in noisy data, e.g. it can detect random permutations of words in a
text \cite{lande}.

\section{The Kolmogorov's stochasticity parameter}

Now we will recall that non-random structures in strings can be also
efficiently detected by means of the Kolmogorov's stochasticity
parameter. This parameter is defined for $n$ independent, ordered,
real-valued random variables $X_1\le X_2\le\dots\le X_n$.  Assuming the
cumulative distribution function of $X$,
\begin{eqnarray}
  \label{eq:7}
F(x) = P\{X\le x\},
\end{eqnarray}
one defines an {\it empirical distribution function} $F_n(x)$
\begin{eqnarray*}
F_n(x)=
\begin{cases}
0\ , & x<X_1\ ;\\
k/n\ , & X_k\le x<X_{k+1},\ \ k=1,2,\dots,n-1\ ;\\
1\ , & X_n\le x\ .
\end{cases}
\end{eqnarray*}
The Kolmogorov's stochasticity parameter $\lambda_n$ is defined as
\cite{Kolm,Arn,Arn_UMN,Arn_MMS}
\begin{equation}\label{KSP}
\lambda_n=\sqrt{n}\ \sup_x|F_n(x)-F(x)|\ .
\end{equation}
Kolmogorov proved \cite{Kolm} that for any continuous cumulative
distribution function $F$:
\begin{eqnarray}
  \label{eq:6}
\lim_{n\to\infty}P\{\lambda_n\le\lambda\}=\Phi(\lambda)\ ,  
\end{eqnarray}
where the limit converges uniformly, and where $\Phi(\lambda)$ is the
fourth elliptic function
\begin{eqnarray}
\label{Phi}
\Phi(\lambda>0)=\vartheta_4(0, e^{-\lambda^2})=
\sum_{k=-\infty}^{+\infty}\ (-1)^k\ e^{-2k^2\lambda^2},
\end{eqnarray}
where $\Phi(0)=0$.
Now $\Phi(\lambda)$ is invariant with respect to $F$, hence it is
universal. As a function of $\lambda$, $\Phi(\lambda)$ monotonously
grows from $0$ to $1$. It is close to zero for $\lambda<0.4$, where
$\Phi(0.4)=0.0028$, and close to $1$ for $\lambda>1.8$, where
$\Phi(1.8)=0.9969$.

Eqs.~(\ref{eq:7}--\ref{Phi}) is the base for the standard
Kolmgorov-Smirnov test. A simpler implementation of this test was
proposed by Arnold \cite{Arn_UMN}. Let $\lambda^*$ be the empirically
observed value of $\lambda_n$. It was shown that even for a {\it
  single} bit string, $\Phi(\lambda^*)\approx 0$, e.g. $\lambda^*
<0.4$, is a good indicator of randomness \cite{Arn_UMN,Arn_MMS}.

To illustrate the power of this method in revealing the degree of
randomness of sequences, we will consider the following example. Let us
consider sequences as superposition of random and regular
sequences. Namely, we will consider one-parameter sequences defined as
\begin{equation}
z_n = \alpha x_n + (1-\alpha) y_n,
\end{equation}
where $x_n$ are random sequences and 
\begin{equation}
y_n = \frac{a\, n \,{\rm mod}\, b}{b},
\end{equation}
are regular sequences defined within the interval $(0,1)$, and 
$a$ and $b$ are mutually fixed prime numbers, parameter $\alpha$ 
while varying within $0$ and $1$ is indicating the variation in 
the fraction of random and regular sequences.

Thus we have $z_n$ with a cumulative distribution function 

\begin{equation}
F(X)= \left\{
\begin{array}{rl}
	0, & X \leq 0 \\
	\frac{X^2}{2 \alpha (1-\alpha)}, & 0 < X \leq \alpha\\
	\frac{2 \alpha X - \alpha^2}{2 \alpha (1-\alpha)}, & \alpha < X \leq 1-\alpha\\
	1-\frac{(1-X)^2}{2 \alpha (1-\alpha)}, & 1-\alpha < X \leq 1\\
	1, & X > 1.\\
\end{array}
\right.
\label{eq:d}
\end{equation}

We will inquire into the stochastic properties of $z_n$ vs the
parameter $\alpha$ varying within the interval $[0,1$ for fixed $a$
and $b$, in order to follow the transition of the sequences from
regular to random ones.

\begin{figure}[!htbp]
  \centering
  \includegraphics[width=120mm]{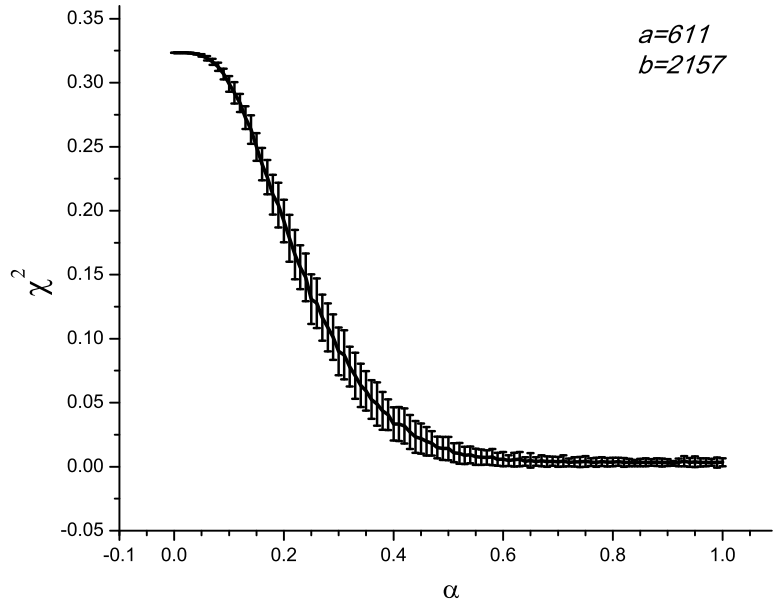}
  \caption{The $\chi^2$ for the sequence $z_n$ vs the parameter
    $\alpha$ reflecting the variation of the ratio of random and
    regular fluctuations.\label{fig:chi_sq}}
\end{figure}

For each value of $\alpha$ for fixed $a$ and $b$, namely, $a=611$,
$b=2157$, 100 sequences $z_n$ were generated, of 10,000 elements the
each sequence. The latter then were divided into 50 subsequences and
for each subsequence parameter $\Phi(\lambda_n)_m$ was obtained and
hence the empirical distribution function $G(\Phi)_m$ of them was
determined ($m$ varied within $1, ..., 50$). In accord to Kolmogorov's
theorem, for example, for purely random sequences, the empirical
distribution should be uniform. So, we estimated $\chi^2$ of the
functions $G(\Phi)_m$ and $G_0(\Phi)=\Phi$ as an indicator for
randomness. Grouping $100$ $\chi^2$ values per one value of $\alpha$,
we constructed mean and error values for $\chi^2$.  The
dependence of $\chi^2$ vs $\alpha$ for given values of $a$ and $b$ is
shown in Fig.\ref{fig:chi_sq}, revealing the informativity of
Kolmogorov's parameter approach in quantitying the degree of non-randomness 
in the sequences.

The Kolmogorov's parameter has been applied, for example, in revealing
of non-random structures in genomic sequences \cite{11}. Here is a
sample piece of the latter:
\begin{eqnarray}
  \label{eq:8}
ACAGAGCTGAGTCACGTGGTGGAAT...,  
\end{eqnarray}
where $A$, $G$, $C$ and $T$ stand for adenine, guanine, cytosine, and
thymine, respectively. Such non-random structures were employed for
detecting mutations in human genome sequences, because mutations can
be related to their regions with locally higher randomness \cite{11}. In
particular, it was shown that only 3-base nucleotide (codon) coding
enables to distinguish somatic mutations.

\section{Zipf's law}

It is natural to look for general features of communication systems,
those which will presumably be present in meaning-conveying
(non-random) messages, even if it is not (yet) known how decipher this
meaning \cite{mccowan}. One general feature of man-created texts|that
remains stable across different languages, and to some extent can be
traced out as well in certain animal communication systems
\cite{mccowan}|is the law discovered by Estoup \cite{est} and
independently (and much later) by Zipf \cite{zipf}; see
\cite{mandelbrot,sole,mitra,manin,arapov,simon,zane,kanter} for an
incomplete list of references and \cite{baa} for a review. This Zipf's
law states that ordered frequencies
\begin{eqnarray}
  \label{eq:1}
f_1\geq f_2\geq ... \geq f_m,  
\end{eqnarray}
of words display a univeral dependency 
\begin{eqnarray}
  \label{eq:5}
f_r\propto r^{-1}  
\end{eqnarray}
of the frequency on the rank $r$, $1\leq r\leq m$. The law expressed
by (\ref{eq:5}) gets modified for very rare words, those with
frequency ${\cal O}(\frac{1}{m})$. This set of rare words is known as
{\it hapax legomena}. However, the Zipf's law efficiently generalizes
to this situation, so that a single expression diescribes well both
moderate and small frequencies \cite{armen}.

The main point of looking at such rank-frequency relation is that once
they hold the Zipf's law, this fact will likely indicate on a
non-random structure implicit in a message, and allow to identify this
message as a text \cite{mccowan,elliot}.

The precise origin of the Zipf's law is not yet completely clear. Zipf
himself believed that this origin is to be sought in specific
co-adaptation of the speaker (coder) and hearer (decoder) \cite{zipf}:
the former finds it economical to employ a possibly smaller set of
words for denoting possibly larger set of meanings. Hence the speaker
tends to minimize the redundancy.  In contrast the hearer (decoder)
would like to introduce sufficient redundancy, so as to minimize
communication errors. The qualitative idea by Zipf was formalized in
several derivations (of certain aspects) of the Zipf's law
\cite{mandelbrot,mitra,arapov}. However, these derivations do not
explain all the relevant aspects of the law, e.g. its unique extension
to the hapax legomena domain. Such aspects are explained by the
derivation of the law proposed in \cite{armen} that deduces the law
from rather general properties of the mental lexicon, the cognitive
(mental) system responsible for the organization of speech and thought
\cite{armen}.

The Zipf's law gets modified in a definite way if one goes out of the
realm of alphabetic languages. E.g. in a sufficiently long Chinese
text the frequencies of characters (that are more morphemes than
words) follow a modified Zipf's law \cite{deng}. However, these
modifications concern only relatively low-frequency characters
\cite{deng}. Moreover, for sufficiently short Chinese texts the Zipf's
law holds fully and is not distinguishable from its standard form of
English texts \cite{deng,armen}.

\section{Conclusions}

We expanded over the information panspermia hypothesis proposed in
\cite{G} and reviewed in \cite{webb}. The basic implication of this
hypothesis is that alien messages may not be rare, but they are
compressed. Recognizing such messages may be challenging, since they
may look like a noise. In fact, an ideally compressed message should
be operationally (algorithmically) random, according to the
Kolmogorov's complexity theory; see section \ref{ks} for a
remainder. However, this ideal compression is not achievable in
principle, because it relates to the undecidability. Thus practically
compressed messages can still show certain ordered patterns.  In this
contribution we considered several methods that can detect those
patterns. 

Though the ideal compression is not achievable in principle, there are
several optimal and universal compression algorithms. Here optimal
means that for a long random data the algorithm compresses according to
the first Shannon's theorem. Following, the methodology proposed by
Minsky \cite{minsky}, one can conjecture that some of those
compression schemes is known to aliens. Here we analyzed one of the
most widespread schemes of this kind, the Lempel-Ziv-Welch compression
algorithm.

Next, we considered the Kolmogorov's stochasticity parameter, an
approach closely related to the Kolmogorov-Smirnov test
\cite{Arn,Arn_UMN,Arn_MMS}. This parameter also has a substantial
degree of universality and was recently shown to be very effective in
detecting genetic mutations \cite{11}. Searching for mutations can be
regarded as looking for locally random domains in DNA \cite{11}.

Finally, we turned to the Zipf's law: (again) a universal law that
folds for ordered word frequencies of (almost all) human texts. The
law does indicate on a non-random structure|though the underlying
cause of this structure is not yet clear|and it can survive after
compression.

The common thread of these methods is that they are
universal, i.e. independent from details of data. This is a
principal condition in searching for extra-terrestrial messages.


\begin{thebibliography}{}

\bibitem{webb} S. Webb, {\it If the Universe Is Teeming with Aliens
    ... Where is Everybody?: Seventy-Five Solutions to the Fermi
    Paradox and the Problem of Extraterrestrial Life} (Springer, NY
  2015).

\bibitem{G}V.G. Gurzadyan, Observatory, {\bf 125}, 352 (2005).

\bibitem{K}A.N. Kolmogorov, Probl. Inform. Transfer, {\bf 1}, 3
  (1965). 

\bibitem{cover}T. Cover and J. Thomas, {\it Elements of Information
    Theory} (Wiley, New York, 1991).
    
\bibitem{minsky}M. Minsky, {\it Why intelligent aliens will be
  intelligible},  Extraterrestrials: Science and Alien
  Intelligence, Vol. 1 (1985).

\bibitem{shostak}S. Shostak, Progress
  in the Search for Extraterrestrial Life, {\bf 74}, 447 (1995).


\bibitem{martin} P. Martin-Lof, Information and Control, {\bf 9},
  602, (1966).

\bibitem{est} J.B. Estoup, {\it Gammes st\'enographique} (Institut
  St\'enographique de France, Paris, 1916).

\bibitem{zipf}G.K. Zipf, {\it The Psycho-Biology of Language: An
    Introduction to Dynamic Philology} (Cambridge, MA, MIT Press,
  1965).

\bibitem{baa}H. Baayen, {\it Word frequency distribution} (Kluwer
  Academic Publishers, 2001).

\bibitem{mccowan}L.R. Doyle, B. McCowan, S. Johnston and S.F. Hanser,
  Acta Astronautica, {\bf 68}, 406-417 (2011).


\bibitem{mandelbrot} B. Mandelbrot, {\it Fractal geometry of nature}
  (W. H. Freeman, New York, 1983).


\bibitem{sole}R. Ferrer-i-Cancho and R. Sol\'e, PNAS, {\bf 100}, 788 (2003).

\bibitem{mitra} B. Corominas-Murtra, J. Fortuny and R.V. Sole,
  Phys. Rev. E {\bf 83}, 036115 (2011).

\bibitem{manin} D. Manin, Cognitive Science, {\bf 32}, 1075 (2008).

\bibitem{arapov} M.V. Arapov and Yu.A. Shrejder, in {\it Semiotics
    and Informatics}, v. 10, p. 74 (Moscow, VINITI, 1978).


\bibitem{simon} H.A. Simon, Biometrika {\bf 42}, 425 (1955).

\bibitem{zane}
D.H. Zanette and M.A. Montemurro, J. Quant. Ling. {\bf 12}, 29
(2005).

\bibitem{kanter}
I. Kanter and D.A. Kessler, Phys. Rev. Lett. {\bf 74}, 4559
(1995).


\bibitem{elliot} J. Elliott, 
Acta Astronautica, {\bf 78}, 26-30 (2012).


\bibitem{Kolm}
A.N. Kolmogorov, Giorn.Ist.Ital.Attuari, 4, 83 (1933)

\bibitem{Arn}
V. Arnold, Nonlinearity, 21, T109 (2008) 

\bibitem{Arn_UMN} V.I. Arnold, Uspekhi Mat. Nauk, 63, 5 (2008) 

\bibitem{Arn_MMS} V.I. Arnold, Trans. Moscow Math. Soc., 70, 31 (2009) 


\bibitem{11} V.G. Gurzadyan, H. Yan, G. Vlahovic, et al, Royal Soc. Open Sci. {\bf 2}, 150143 (2015).


\bibitem{armen} A. E. Allahverdyan, W. Deng and Q. A. Wang,
  Phys. Rev. E {\bf 88}, 062804 (2013).

\bibitem{deng}  W. Deng, A. E. Allahverdyan, Bo Li and Q. A. Wang,
Eur. Phys. J. B, {\bf 87}, 47 (2014).

\bibitem{lz}
J. Ziv and A. Lempel, 
IEEE Transactions on Information Theory, {\bf 23}, 337--342 (1977).

J. Ziv and A. Lempel, 
IEEE Transactions on Information Theory, {\bf 24}, 530--536 (1978).

T. A. Welch, Computer, 8--18 (1984).

\bibitem{lz_complexity} A. Lempel, J. Ziv, IEEE Transactions on
  Information Theory, {\bf 22}, 75-81 (1976)

\bibitem{aboy}M. Aboy, R. Hornero, D. Abasolo and D. Alvarez, IEEE
  Trans. Biomed. Eng. {\bf 53}, 2282-2288 (2006).


\bibitem{lande} D.V. Lande, A.A. Snarskii, {\it On the role of
    autocorrelations in texts}, arXiv:0710.0225. 



\end{thebibliography}
\end{document}